# Observation of plasmon-phonon coupling in natural 2D graphene-talc heterostructures

Ingrid D. Barcelos,[a]* Alisson R. Cadore,[b]* Ananias B. Alencar,[c] Francisco C. B. Maia,[a] Edrian Mania,[b] Rafael F. Oliveira,[d] Carlos C. B. Buffon,[d] Ângelo Malachias,[b] Raul O. Freitas,[a] Roberto L. Moreira,[b] Hélio Chacham[b]

[a]Laboratório Nacional de Luz Síncroton (LNLS), Centro Nacional de Pesquisa em Energia e Materiais (CNPEM) 13083-970, Campinas, São Paulo, Brazil
[b]Departamento de Física, Universidade Federal de Minas Gerais (UFMG) 30123-970, Belo Horizonte, Minas Gerais, Brazil
[c]Instituto de Engenharia, Ciência e Tecnologia Universidade Federal dos Vales do Jequitinhonha e Mucuri (UFVJM) 39440-000, Janaúba, Minas Gerais, Brazil
[d]Laboratório Nacional de Nanotecnologia (LNNano), Centro Nacional de Pesquisa em Energia e Materiais (CNPEM) 13083-970, Campinas, São Paulo, Brazil
*These authors contributed equally to this work.
*Corresponding-Author*: ingrid.barcelos@lnls.br

Two-dimensional (2D) materials occupy noteworthy place in nanophotonics providing for subwavelength light confinement and optical phenomena dissimilar to those of their bulk counterparts. In the mid-infrared, graphene-based heterostructures and van der Waals crystals of hexagonal boron nitride (hBN) overwhelmingly concentrate the attention by exhibiting real-space nano-optics from plasmons, phonon-polaritons and hybrid plasmon phonon-polaritons quasiparticles. Here we present the mid-infrared nanophotonics of talc, a natural atomically flat layered material, and graphene-talc (G-talc) heterostructures using broadband synchrotron infrared nano-spectroscopy. We achieve wavelength tuning of the talc resonances, assigned to in- and out-of-plane vibrations by changing the thickness of the crystals, which serves as its infrared fingerprints. Moreover, we encounter coupling of the graphene plasmons polaritons with surface optical phonons of talc. As in the case of the G-hBN heterostructures, this coupling configures hybrid surface plasmon phonon-polariton modes causing 30 % increase in intensity for the out-of-plane mode, blue-shift for the in-plane mode and we have succeeded in altering the amplitude of such hybridization by varying the gate voltage. Therefore, our results promote talc and G-talc heterostructures as appealing materials for nanophotonics, like hBN and G-hBN, with potential applications for controllably manipulating infrared electromagnetic radiation at the subdiffraction scale.

## Introduction

Talc is a natural soft magnesium silicate mineral with a crystalline structure that contains three octahedral Mg positions per four tetrahedral Si positions with the chemical formula[1–3] $Mg_3Si_4O_{10}(OH)_2$. In addition to hydrophobicity and high thermal stability, talc nanocrystals are chemically inert and possess atomically flat surfaces.[4–6] Similarly to other two-dimensional (2D) materials, e.g., molybdenum disulfide, hexagonal boron nitride (hBN) and graphene, talc is formed by a lamellar structure held together by van der Waals forces.[7] Recently, outstanding 2D elastic modulus of mechanically exfoliated few layers talc was shown to be comparable to the graphene one.[8] The bandgap (~5.3 eV) of talc directly compete with the well-established $SiO_2$ insulator. In contrast to $SiO_2$, talc is a genuine 2D material whose attributes match its use as insulating media compatible with atomically thin P-N junctions and field effect transistors for ultra-compact devices.[9,10] As a matter of fact, talc induces high *p*-type doping in graphene, preserves its high electronic mobility, and allows the observation of quantum Hall effect in graphene at low magnetic fields.[11] These mechanical and electronic virtues are augmented by the strong optical activity of talc in the infrared range, an important aspect generated by its polar crystal structure still barely exploited.

As hBN, talc nanocrystals bear possibility of controlling subdiffractional light. Here, we systematically investigate the opto-vibrational properties of isolated talc nanocrystals and graphene-talc (G-talc) heterostructures on different substrates (metallic, gold and dielectric, $SiO_2$), using synchrotron ultra-broadband infrared nano-spectroscopy (SINS).[12–14] As we present, the infrared optical responses of graphene and talc combine to make the G-talc heterostructure a nanophotonic electromagnetic hybrid.[15,16] We report three resonant surface phonons polaritons in the SINS spectra of talc nanocrystals. Through first-principles calculations, the resonance near to 670 $cm^{-1}$ is assigned to the in-plane librations of hydroxyl groups, i.e, with the resulting infrared dipole parallel to the basal plane of the atomic layers. Two resonances near 960 $cm^{-1}$ and 1025 $cm^{-1}$ are, respectively, attributed to in- and out-of-plane stretching modes of the Si-O bonds. It is experimentally noticed that the resonance frequencies can be tuned as a function of the number of atomic layers, and can be readily served as its infrared fingerprints. Furthermore, we show that single graphene layer atop the talc nanocrystals leads to important modifications to the in- and out-of-plane Si-O stretching modes: an increase in the intensity (> 30 %) for the out-of-plane mode at 1025 $cm^{-1}$ and as a 30 $cm^{-1}$ blue-shift for the in-plane mode at 960 $cm^{-1}$. Finally, we also demonstrate that the primer characteristic can be systematically tuned by varying the gate voltage, evidencing the observation of the coupling of graphene plasmon polaritons with surface phonons polaritons of talc stemming the surface plasmon phonon-polariton analogous to the reported for graphene-$SiO_2$ [17–19] and graphene-hBN[19–24] heterostructures. Hence, natural 2D talc nanocrystals raise as a potential nanophotonic material with configurable infrared activity.[15,25–28]

## Experimental Methods

*General characterization of talc nanocrystals and fabrication of G-talc heterostructures:* A talc crystal, obtained from Ouro Preto (MG), Brazil (talc/soapstone mine), was initially characterized by Synchrotron X-Ray Diffraction and Fourier Transform Infrared (FTIR) Spectroscopy (see Supporting Information). Conventional polarized-reflectivity and absorption FTIR spectra of bulk samples (Thermo-Nicolet equipment, Centaurus microscope and Mercury Cadmium Telluride (MCT) detectors) were useful for characterizing the observed features in the SINS spectra. Standard graphene scotch tape exfoliation method was carried out to release and transfer mono- and multi-layers of talc atop two different substrates: $SiO_2$/Si (285 nm thick $SiO_2$) and a thermally evaporated Au film (100 nm thick) on bare Si. The resulting flakes were pre-characterized by optical microscopy and by atomic force microscopy (AFM) to determine the number of layers.[8] The G-talc/$SiO_2$ (Figures 1c and 1f) and G-talc/Au (Figures 1d and 1g) heterostructures were assembled by transference of monolayer graphene to the top of talc flakes previously deposited on the respective substrates. This procedure is the same used for the fabrication of graphene-hBN heterostructures.[29,30] The samples were then submitted to heat cleaning process at 350 Celsius, with constant flow of Ar/$H_2$ (300:700 sccm) for 4 h, to remove organic residue from the transferring process. The exfoliated graphene was pre-characterized by optical microscopy and Raman spectroscopy to confirm that it consisted of monolayer graphene.[31] We studied G-talc/Au samples with talc thickness ranging from 30 nm to 100 nm, and all of which produced consistent results.

*Electrical characterization and gate-tunability of G-talc/Au devices:* We fabricated a G-talc heterostructure atop to a stripe of gold designed by standard electron-beam lithography and thermal metal deposition (40 nm thick) on $SiO_2$/Si substrate. Initially, talc flake was transferred atop of the gold stripe, followed by the single layer graphene forming the G-talc/Au heterostructures. Finally, the metallic electrodes were defined by a second electron-beam lithography and thermal deposition (60 nm Au thick). The Figure 6a depicts the optical image of one ordinary G-talc/Au device measured in this work. For the measurements as a function of the back-gate voltage ($V_G$), the bottom gold stripe was used as the back-gate electrode, and we worked with safe limits of gate bias, avoiding leakage current through the dielectric. Moreover, the electrical characterization of the G-talc/Au transistor was carried out in $N_2$ atmosphere (humidity levels down to 2-3 %) at room temperature using a Keithley 2636 Source-Meter® unit. The source-drain current ($I_{SD}$) was acquired by sweeping the gate voltage from 0 V to 35 V, at 250 mV/s, with a constant source-drain voltage $V_{SD} = 1$ mV.

*Near-Field Optical Microscopy Measurements*: The SINS experiment consists of scattering-type scanning near-field optical microscopy (s-SNOM),[12–14,32,33] using a commercial microscope (NeaSNOM, Neaspec GmbH), wherein the broadband infrared (IR) beam, delivered by the bending magnet beamline of the Brazilian Synchrotron Light Laboratory (LNLS, Campinas), is the excitation light source.[34] In brief, the IR radiation induces the antenna effect to a metal-coated AFM tip (ARROW-NCPt-W, Nano World) of the s-SNOM microscope. High intense optical fields confined at the tip apex cause an effective polarization to the tip-sample system. The optical near-field corresponds to light scattered from the polarized tip-sample system. This radiation originates from an area on the sample defined by the tip apex, typically, an area of ~ 25 nm diameter. Typically, in s-SNOM, the tip-sample scattered light is modulated by the mechanical frequency of the tip ($\Omega \approx 300$ kHz, in our case) since AFM operates in semi-contact mode. The background-free optical near-field is given by the high harmonics ($n\Omega$, with $n \geq 2$) of the tip-sample scattered radiation, which are detected by a lock-in based scheme using a MCT detector (IRA-20-00103, Infrared Associates Inc.). Here, our data refer to the $2\Omega$ optical near-field. Concomitantly, AFM topography and spectrally integrated optical near-field IR images are obtained. To realize the synchrotron infrared nano-spectroscopy, the optical near-field carrying broadband response from the tip-sample interaction is sent to an asymmetric Michelson interferometer that permits retrieving, by Fourier Transform, the amplitude and phase spectra. This setup allows for analysis via point-spectrum and spectral linescan. In our measurements, the point-spectrum consists of an average of 20 spectral scans, each scan divided in 2048 points with 20 ms integration time per point. The spectral linescan corresponds to a set of spectra acquired along a defined distance (line) yielding the spectral map (wavenumber $\times$ distance $\times$ intensity), as shown in Figure 5a. All SINS spectra here are normalized to the reference spectrum from a non-resonant Au surface.

**Results and discussion**

Figure 1a presents, schematically, the SINS experiment for the G-talc heterostructures and talc nanocrystals. As illustrated, the IR beam confined at the tip apex excites the hybrid surface plasmon phonon-polariton ($SP^3$) modes, which will be explained herein, on the G-talc surface. In the Figures 1b-d, we show AFM topographies of multilayer flakes of talc on $SiO_2$ substrate, G-talc/$SiO_2$ and G-talc/Au heterostructures, respectively. These measurements feature atomic layer steps in multiples of 1 nm, corresponding to the thickness of an exfoliated monolayer of talc.[8] Sharp optical contrast for different number of layers, graphene and the substrate are noticed, respectively, in near-field images of the Figures 1e-g.

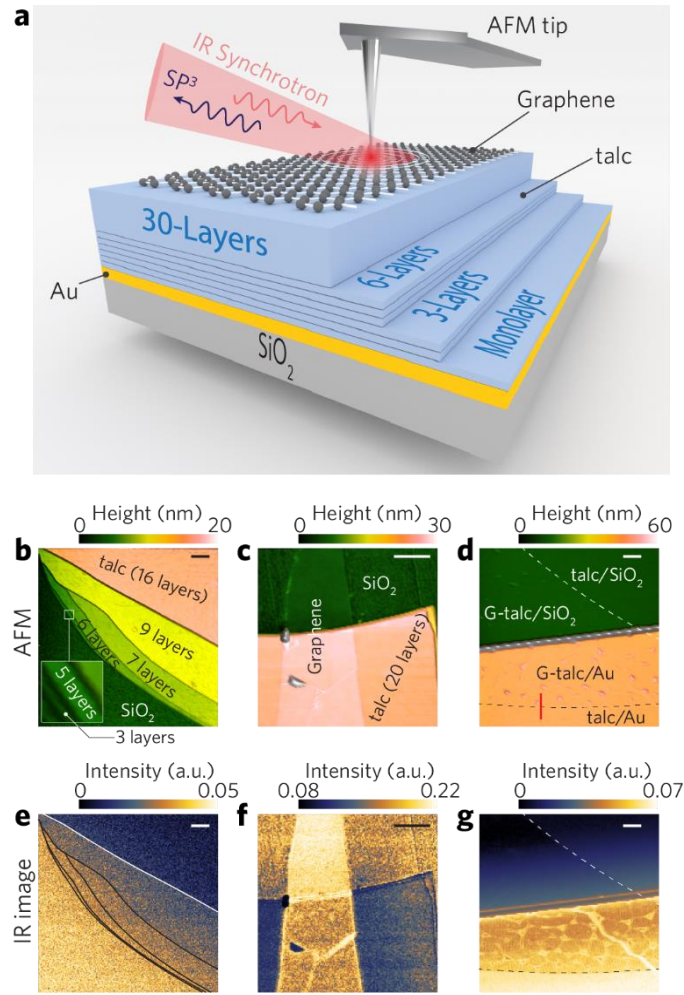

**Figure 1:** Near-field analyzes of G-talc heterostructures. (a) Schematic of SINS experiment, G-talc heterostructure and talc nanocrystals. In the tip-sample interaction, the IR confined radiation excites $SP^3$ modes on the G-talc heterostructure. AFM topography: (b) multi-layer flake of talc on a $SiO_2$ substrate, (c) G-talc/$SiO_2$ heterostructure and (d) G-talc/Au heterostructure. Near-field images of the same regions of the topographies are shown in (e), (f) and (g), revealing scattering intensity modulation according to the number of talc layers. Dashed lines mark the edges of the single layer graphene in (d) and (g). Solid lines in (e) are guides for the talc layers' boundaries shown in (b). The red line marked in (d) depicts the region of the spectral linescan across the talc/Au and the graphene/talc/Au border shown in Figure 5a. In all figures, the scale bar represents 1 μm.

The spectral response of the talc layers is further explored by SINS. In the Figure 2a, the point-spectra of talc nanocrystals on Au substrate with different number of layers present three resonances. The referred peaks are related to IR active vibrations of bulk talc crystals[2,3,7,35,36] and, herein, are ascribed to in- and out-of-plane vibrational modes in agreement with our calculations detailed in the following. For the 6 nm thick crystal, the maxima of those peaks appear near to 680, 990 and 1010 cm$^{-1}$. Such resonances, however, differ in lineshape, intensity and spectral position as a function of the thickness increase as shown in Figure 2b.

These results indicate that those talc modes can be spectrally tuned upon controlling the number of layers, and readily determine the fingerprints of different number of layers. Note that similar results are also observed for the G-talc/SiO$_2$ heterostructure (see Supporting Information).

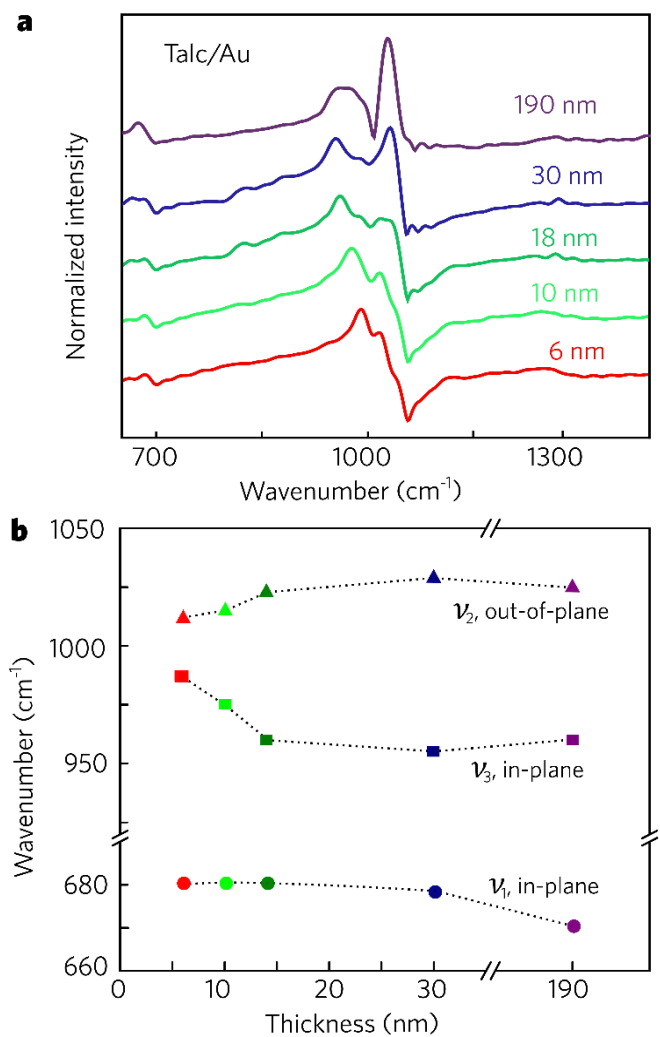

**Figure 2:** Infrared fingerprints of talc layers. (a) SINS point-spectra of talc nanocrystals with different number of layers on the Au substrate. The spectra are dominated by the silicate IR active vibrations. (b) Position of the maxima of the talc peaks, which are labeled according to theoretical assignment, as a function of the talc thickness on the gold substrate.

The mode assignments are performed by comparing experimental observations with first-principles calculations considering two limit cases: bulk and single-layer talc. The calculations employed the density functional theory (DFT),[37,38] and the density functional perturbation theory (DFPT)[39] as implemented in Plane-Wave Self-Consistent Field (Pwscf) and Phonon packages of the Quantum-Espresso distribution.[40] The plane-wave kinetic energy cutoff to describe the electronic wave functions was set to 100 Ry. The Brillouin zone integrations are made within the Monkhorst-Pack scheme[41] by using 3x3x2 and 3x3x1 grids for bulk and monolayer talc. We used the triclinic P$\bar{1}$ unity cell for the bulk talc.[42] The monolayer talc was treated within the slab model, with images separated by 30 Å of vacuum. The convergences for the lattice and atomic position optimizations were achieved for stress lower than 0.1 kBar and forces lower than 0.1 mRy/Bohr. We used the PBE GGA functional[43] rigorously double-checked for negative frequencies using the LDA functional. The Ionic cores are represented by norm-conserving pseudopotentials.[44]

Figure 3a shows the calculated IR absorption spectra (ε" optical function) of both bulk and monolayer talc, from the DFT calculations. The complete dispersion parameters obtained by DFT necessary for obtaining these spectra are given in the Supporting Information, besides sketches of the atomic movements for the most important modes in our observation region, allowing us to check the origin of such modes. The bulk talc spectrum is similar to the theoretical one in the ref. 3. The simulated spectra are mainly characterized by three peaks indicated as $v_1$, $v_2$ and $v_3$ for bulk and for the monolayer talc. We plot in Figures 3b-d, superimposed with the atomic positions of the unit cell, the relative magnitude and direction (represented by the blue arrows) of the displacements of the talc atoms for the normal vibrational modes with the largest contributions to the calculated peaks shown in the Figure 3a. The orientation of the unit cell in Figure 3b-d is such that the horizontal direction is parallel to the basal plane of talc (the plane that contains the talc layers) and the vertical direction is perpendicular to the basal plane. Thus, we conclude that for both bulk and monolayer talc, $v_1$ corresponds mainly to in-plane librational modes of the hydroxyl groups, $v_3$ corresponds to in-plane Si-O stretching modes, and $v_2$ corresponds to out-of-plane Si-O stretching modes. The calculated in-plane vibrations occur at nearly the same frequency for bulk and monolayer talc systems (around 670 cm$^{-1}$ and 970 cm$^{-1}$, for ($v_1$) and ($v_3$), respectively). In contrast, the out-of-plane mode is calculated at ~ 912 cm$^{-1}$ for the former, and at ~ 1000 cm$^{-1}$ for the latter. Therefore, based on the calculations, we can assign the experimental resonance at 670 cm$^{-1}$ ($v_1$) to the in-plane librational motion of the hydroxyl groups, and the experimental resonances at 960 cm$^{-1}$ ($v_3$) and 1025 cm$^{-1}$ ($v_2$) to Si-O stretching modes. For more details, see Supporting Information.

The frequency proximity of the Si-O stretching modes ($v_2$ and $v_3$), whose difference is of the same order of magnitude as the DFT accuracy, combined with the non-trivial light-matter interaction at the tip-sample system, difficult the identification of the orientation of these modes (in- or out-of-plane). However, with a comprehensive analysis of DFT, SINS, absorption and polarized reflectivity FTIR data presented in the Supporting Information, confirm the proposed attribution is done in the precedent paragraph: $v_3$ is an in-plane mode appearing around 960 cm$^{-1}$ in the SINS spectra, $v_2$ an out-of-plane mode appearing around 1025 cm$^{-1}$.

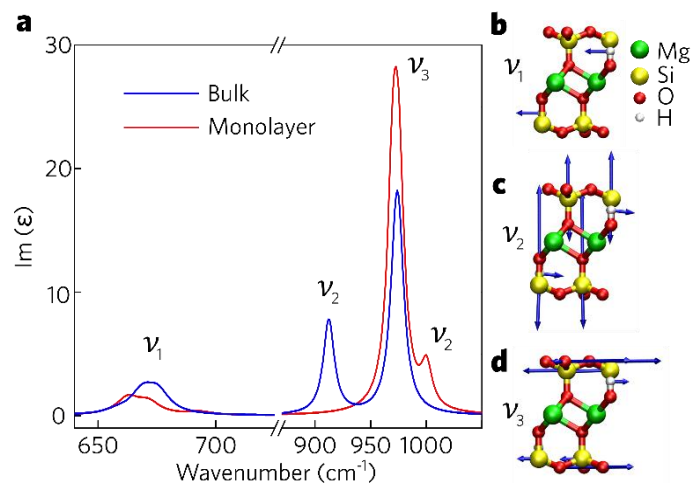

**Figure 3:** Simulated talc modes. (a) Simulated Im($\varepsilon$) functions for bulk and monolayer talc. Relative magnitude and direction (blue arrows) of the displacements of talc atoms in the normal vibrational modes with the largest contribution to peaks $\upsilon_1$, $\upsilon_2$, and $\upsilon_3$ of bulk and monolayer talc shown in (b), (c) and (d). The horizontal direction is parallel to the talc basal plane. Therefore, modes $\upsilon_1$ and $\upsilon_3$ are predominantly in-plane modes, and mode $\upsilon_2$ is predominantly out-of-plane. Si, Mg, O and H atoms are represented by yellow, green, red and white spheres, respectively.

The vibrational response of the talc nanocrystal is significantly modified in the G-talc heterostructure as demonstrated by the spectral linescan in the Figure 4b. From 0 to 600 nm, the spectral linescan probes the G-talc heterostructure and, from 600 to 1000 nm, the talc/Au structure (indicated in Figure 4a). In the talc/Au region, we note the herein discussed modes at 960 cm$^{-1}$ and at 1025 cm$^{-1}$ strictly ascribed to talc. The impact of graphene layer is particularly prominent in the out-of-plane mode (1025 cm$^{-1}$) where the resonance mode is significantly enhanced (> 30 %), with a small blue-shift (6 cm$^{-1}$) and there is also a blue-shift by nearly ~ 30 cm$^{-1}$ for the in-plane mode (960 cm$^{-1}$). The increase in amplitude for the 1025 cm$^{-1}$ mode and the small blue-shift are clearly noticed from the G-talc point-spectrum shown the Figure 4c could be related to the high density of mobile carriers present in our graphene samples.[11] Such phenomenon is equivalent to the coupling effect reported for graphene-hBN heterostructures[19–23] wherein only the hBN out-of-plane mode experiences enhancement. Therefore, in G-talc heterostructures plasmon-polaritons of graphene couple to surface phonons polaritons of talc creating the hybrid SP$^3$ modes in similar way graphene plasmons were previously reported to interact with SiO$_2$ surface phonons.[17–19] Increase of the out-of-plane mode occurs also for the G-talc/SiO$_2$ heterostructure (see Supporting Information), thus, pointing that the SP$^3$ hybridization in G-talc heterostructures happens regardless the substrate nature (metal or insulator).

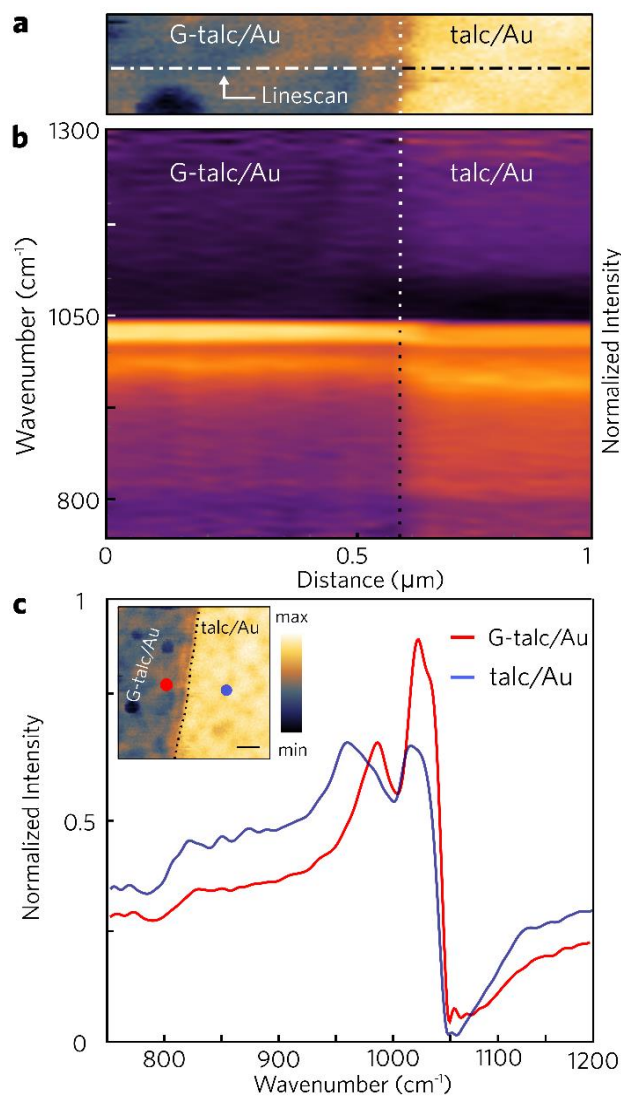

**Figure 4:** Plasmon-phonon coupling in G-talc heterostructures. (a) Near-field image of the sample region wherein the spectral linescan was taken. The vertical dashed lines indicate the transition from the G-talc/Au to the bare talc/Au. (b) Spectral linescan acquired along the blue line indicated in (a). The 1 µm distance was divided in 21 points, with 50 nm step size, with an average of 4 spectral scans per point. (c) Point spectra of the talc/Au and G-talc/Au obtained at the locations marked in the near-field image shown in the inset. The dashed line in the inset delimitates the G-talc/Au and talc/Au regions. The talc thickness is 30 nm. Scale bar represents 500 nm.

We believe that the large blue-shift of ~ 30 cm$^{-1}$ for the $v_3$ mode in the G-talc/Au heterostructure is not clearly observed for the G-talc/SiO$_2$ (see Supporting Information) due to a complex combination of the SiO$_2$ surface phonons at 1120 cm$^{-1}$ and talc modes. Particularly, the in-plane mode at 960 cm$^{-1}$, noticeably seen in talc/Au (see Figure 2), is hardly noticed in talc/SiO$_2$ for similar crystals thicknesses (see Supporting Information). Hence, the blue-shift of this mode is barely distinguishable for the G-talc/SiO$_2$ structure. Additionally, the polarization of the IR excitation and the tip geometry naturally leads to major out-of-plane polarization to the optical near-field. Thus, it is expected a low efficiency for exciting in-plane modes in our SINS experiment. The polarization anisotropy of s-SNOM was recently explored in an experiment of vibrational nano-crystalography.[33] This picture is truly applied for the talc and G-talc on SiO$_2$ and explains the absence of in-plane modes for this substrate. However, an amplification of the optical near-field happens near Au surface due to presence of free charges and formation of a mirror dipole. Accordingly, this high intense optical near-field presents in-plane component with sufficient amplitude to excite the in-plane mode of talc.

Next, according to previous works[17,18,21,45,46] the coupling between the graphene plasmons and the substrate phonons (hBN or SiO$_2$) can be tuned by the charge carrier density. We then demonstrate this aspect of graphene plasmonics experimentally through point spectra under gate bias. Firstly, we present in Figure 6b the transfer curve of one ordinary G-talc/Au device with 100 nm talc thick. The optical image of the device is shown in Figure 5a. As can be seen in Figure 5b, the charge neutrality point of G-talc/Au device is reached by applying back-gate voltages up to $V_G = 32$ V, indicating a high natural $p$-type doping, as expected.[11] Over a range of $V_G$ values from 0 V to 32 V, the hole density in the graphene channel decreases, consequently, this tuning of carrier density produces systematic variations in the G-talc plasmonics profiles as shown in Figure 5c. For instance, there is a significantly enhancement (~ 50 %) in intensity for the out-of-plane resonance mode ($v_2$) for large charge carrier density ($V_G \sim 0\ V$), while the in-plane mode ($v_3$) slightly changes (< 10 %) as a function of the gate bias. Note in the inset of Figure 5c that the impact of the carrier density is clearly prominent in the out-of-plane mode. Therefore, we experimentally demonstrate the gate-tunability of the plasmon-phonon coupling between graphene and talc substrate.

Here, it is important to point out the similarity between our findings for the G-talc heterostructures, with previous works[17,19] for the G-SiO$_2$ nanostructures. For instance, Fei and co-workers[17] observed and modeled the interaction between graphene plasmons and the SiO$_2$ phonons, and demonstrated that such hybridization can be controlled by gate voltage. In such work, the authors found that graphene is able to enhance considerably the amplitude spectra of SiO$_2$, and also blue-shifts the peak frequency by about 10 cm$^{-1}$. Moreover, the authors demonstrated that increasing the hole density in the graphene channel, the spectrum amplitude is further enhanced, and they discussed such phenomena based on plasmon-phonon coupling at the graphene-SiO$_2$ interface. Although more detailed theoretical analysis is needed to fully understand and describe the dispersion relation and loss function of talc and G-talc, we believe that the same phenomenology used to G-SiO$_2$ can be applied for the graphene-talc interface, reinforcing our conclusion that the results obtained in our work can be ascribed by the plasmon-phonon coupling.

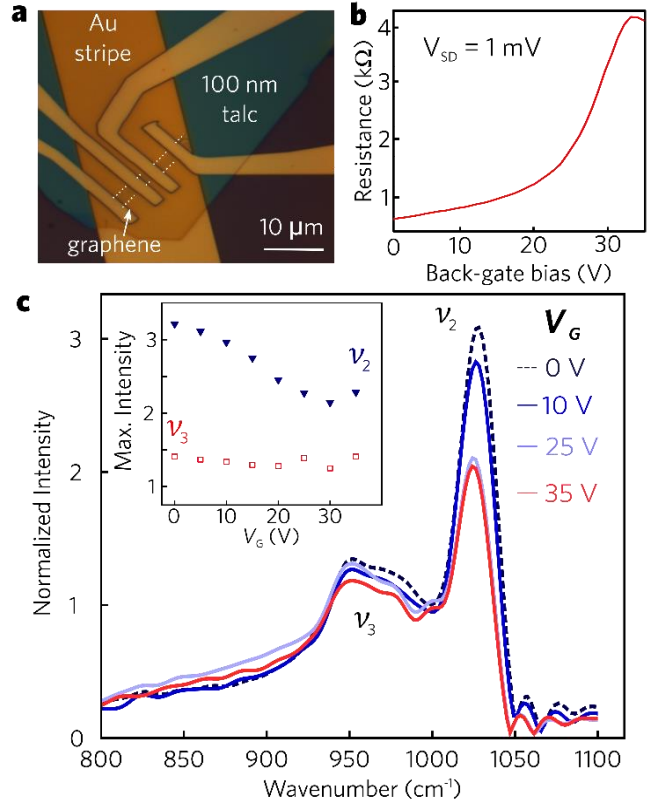

**Figure 5:** Monitoring the resonance spectra with gating voltage. (a) Optical microscope image of one G-talc/Au device using 100 nm talc thick on top of 40 nm Au thick stripe used as back-gate bias. The white dashed lines mark the graphene channel. (b) The resistance as a function of back-gate voltage ($V_G$) for the G-talc/Au heterostructure at room temperature and under N$_2$ atmosphere. (c) Point spectra for the G-talc/Au heterostructure for some selected gate bias, while in the inset are presented the maxima intensities of point spectra for all gating measurements, demonstrating the gate-tunability of the plasmon-phonon coupling between graphene and talc.

## Conclusions

To our knowledge, despite the increasingly prolific research on 2D materials, as the transition metal dichalcogenides, only hBN and talc form ultra-flat layered crystals with optical activity in the technologically attractive mid-infrared range. Hence, talc nanocrystals can potentially reveal intricate photonic properties and these predictions are supported by the observed SP$^3$ hybridization seen in the G-talc heterostructure which has also selection rules for the vibration modes mediated by crystal symmetry. Summarly, we carry out a comprehensive investigation of opto-vibrational properties of talc nanocrystals and the G-talc heterostructures using SINS which is physically interpreted by concise theoretical description of G-talc plasmon-phonon dispersion supported by DFT calculations. The local scattering signal level from near-field IR images allows distinguishing talc nanocrystals with different number of atomic layers, including monolayer. We accomplish the assignment of the vibrational resonances of talc nanocrystals and demonstrate that these frequencies change as a function of the number of layers, which serves as its infrared fingerprints. We also find that the presence of a graphene layer atop the talc nanocrystals leads to considerable modifications in the talc vibrational activity, which were attributed to the coupling of graphene plasmons with surface phonons polaritons of talc regardless the substrate. Despite the fact that analogous coupling is reported to occur in graphene on amorphous $SiO_2$,[17–19] and on hBN substrate,[19–23] we show that the $SP^3$ modes in the G-talc heterostructure affects the in and out-of-plane polarized vibrations of the talc nanocrystals and such hybridization can be tuned by changing the charge density. Thus, we foresee talc as a potential 2D photonic crystal whose vibrational response can be tuned by device fabrication. In addition, the interaction between graphene and talc confirms the capability of the G-talc heterostructure of converting free space radiation into sub-diffractional vibrational activity. Its mechanical robustness, high flatness and natural abundance potentially elects talc and its heterostructures with diverse 2D materials[15,16,27,28] to be a key element in low-cost ultra-compact devices for long-wavelength and polariton-based mid-infrared applications.

## Acknowledgements

This work was supported by CAPES, Fapemig, CNPq and INCT/Nanomateriais de Carbono. R.F.O and C.C.B.B acknowledge financial support from FAPESP (2014/25979-2). The authors thank LNLS for providing the beamtime, Neaspec GmbH for technical support on the s-SNOM hardware at the IR1 beamline of the LNLS, Laboratório de Nanomateriais at UFMG for allowing use of the atomic layer transfer system. The authors also thank the support of the Microfabrication Laboratory at the LNNano for providing substrate metallization, Thiago M. Santos and Vinícius O. da Silva for the technical assistance and Yves Petroff for reading the manuscript.

# Supplementary Information

# Infrared fingerprints of natural 2D talc and plasmon-phonon coupling in graphene-talc heterostructures

Ingrid D. Barcelos,[1,*] Alisson R. Cadore,[2,*] Ananias B. Alencar,[3] Francisco C. B. Maia,[1] Edrian Mania,[2] Ângelo Malachias,[2] Roberto L. Moreira,[2] Raul O. Freitas,[1] Hélio Chacham[2]

[1] Laboratório Nacional de Luz Síncrotron (LNLS), Centro Nacional de Pesquisa em Energia e Materiais (CNPEM), CEP 13083-970, Campinas, São Paulo, Brasil.
[2] Departamento de Física, Universidade Federal de Minas Gerais, 30123-970 - Belo Horizonte, Minas Gerais, Brazil
[3] Instituto de Engenharia, Ciência e Tecnologia, Universidade Federal dos Vales do Jequitinhonha e Mucuri, 39440-000 – Janaúba, Minas Gerais, Brazil
[*] These authors contributed equally.
Electronic address: ingrid.barcelos@lnls.br

CONTENTS:

**Section 1 (S1): X-Ray Diffraction**

**Section 2 (S2): Fourier Transform Infrared Spectroscopy (FTIR)**

**Section 3 (S3): Optical response of Talc/SiO$_2$ heterostructures**

**Section 4 (S4): Graphene-talc plasmon-phonon coupling in SiO$_2$ substrate**

**Section 5 (S5): Optical phonon characteristics from DFT calculations**

**Section 1 (S1): X-Ray Diffraction**

Mineral talc, extracted from a talc/soapstone mine from Ouro Preto (Brazil) was used to produce exfoliated samples that were investigated in this work. The choice of this sample, from which small flakes were exfoliated, was carried out due to its white-green color, which qualitatively indicates a reduced amount of impurities.

In order to corroborate $Mg_3Si_4O_{10}(OH)_2$ as the major phase of the system X-ray diffraction (XRD) measurements were carried out at the XRD1 beamline of the Brazilian Synchrotron Light Laboratory (LNLS). This bending magnet beamline is equipped with a double crystal Si (111) monochromator and a cylindrical Rh mirror, yielding a photon flux of $10^{10}$ photons/sec at the sample position. The measurement system consists of a 24 K mythen strip detector mounted on a Newport heavy-duty diffractometer, allowing for the acquisition of XRD data on a wide angular range. The measurement shown in Figure S1(c) (solid dots) was performed at 12.000 keV (wavelength λ = 1.0332 Å) using a glass capillary with talc powder extracted from the sample shown in Figure S1 (a) and sieved to 10 μm grains. The powder was heated at 150 °C for 2 hours, in order to remove water molecules adsorbed in the system, and promptly measured after quenching to room temperature. The main phase observed is the triclinic system (Rietveld refinement using ICSD # 100682), with lattice structure given by a = 5.291(1) Å, b = 9.172(1) Å, c = 9.455(1) Å, α = 90.58(1) °, β = 98.77(1) °, γ = 89.99(1)°. This crystallographic structure, represented in Figure S1 (b), was used to produce the fit of Figure S1 (c) (Rietveld refinement using the MAUD software), which was the best possible fit using a single talc phase. Other five structures from crystallographic databases were also simulated, leading to poor or incomplete fits of the experimental data. Possible impurities were not simulated (would require an extensive analysis, which is beyond the scientific scope of this work) but may represent an extremely reduced fraction of the atoms in the studied structure.

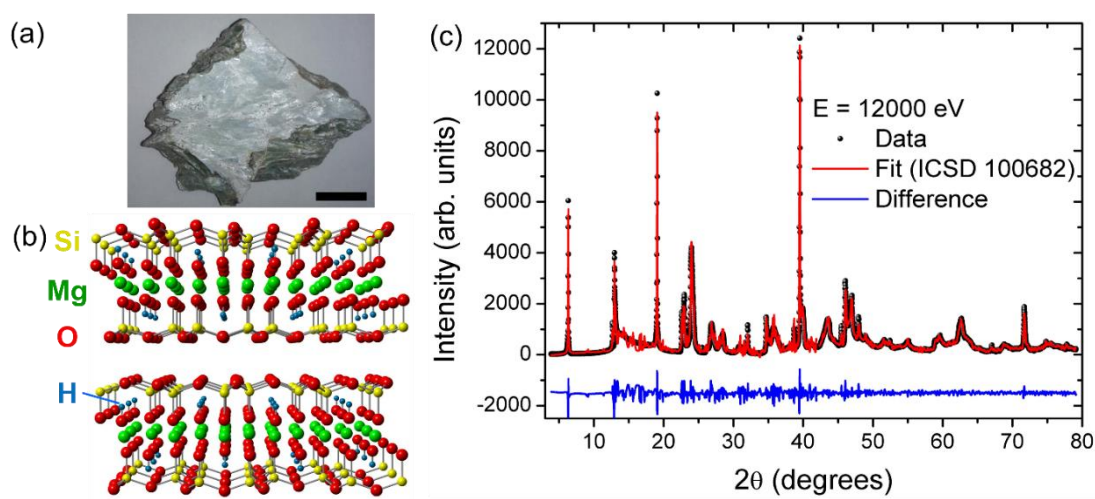

**Figure S1:** (a) Mineral talc obtained from a talc/soapstone mine in Ouro Preto (Brazil). Scale bar is 2.5 cm. (b) Representation of the triclinic crystalline structure of talc (ICSD # 100682), used to fit the diffraction data of (c).

**Section 2 (S2): Fourier Transform Infrared Spectroscopy (FTIR)**

In order to characterize the optical quality of the investigated mineral talc and its infrared polar phonon modes, conventional FTIR spectra were acquired using different experimental approaches. The measurements were performed on a Thermo-Nicolet NEXUS 470 spectrometer, equipped with a Centaurus microscope (10 × magnification, observation region of 250×250 μm$^2$) and MCT detectors. Very nice nearly transparent (greenish) samples were taken from a big polycrystal (10×9×4 cm$^3$) show in Figure S1 (a). Absorption spectra were acquired from transmission through pelletized samples (2.5 mg of powdered talc into 300 mg KBr) in the spectrometer macro-chamber, in the mid infrared range (550-4000 cm$^{-1}$). Polarized and unpolarized reflectivity spectra were collected on oriented flake samples, under microscope, in the same spectral region. Gold mirror was used as reference (backgrounds were collected for each polarization angle). In all cases, 32 scans were averaged and the spectral resolution was better than 4 cm$^{-1}$.

Figure S2 (a) shows that the SINS spectrum of the nanostructured talc/Au present the main characteristic features showed by conventional FTIR spectra of the bulk material either, in absorption or reflection modes. One can observe that the SINS spectrum resembles a reflectivity one, but it is worthy to note that each spectrum is indeed linked to a different optical function – which is related to the same vibrational modes, of course. Also, there is a clear downshift in the frequencies of the two highest frequency modes observed by SINS when compared to conventional bulk measurements. This downshift is certainly a consequence of the scattering process that generates the SINS spectra.

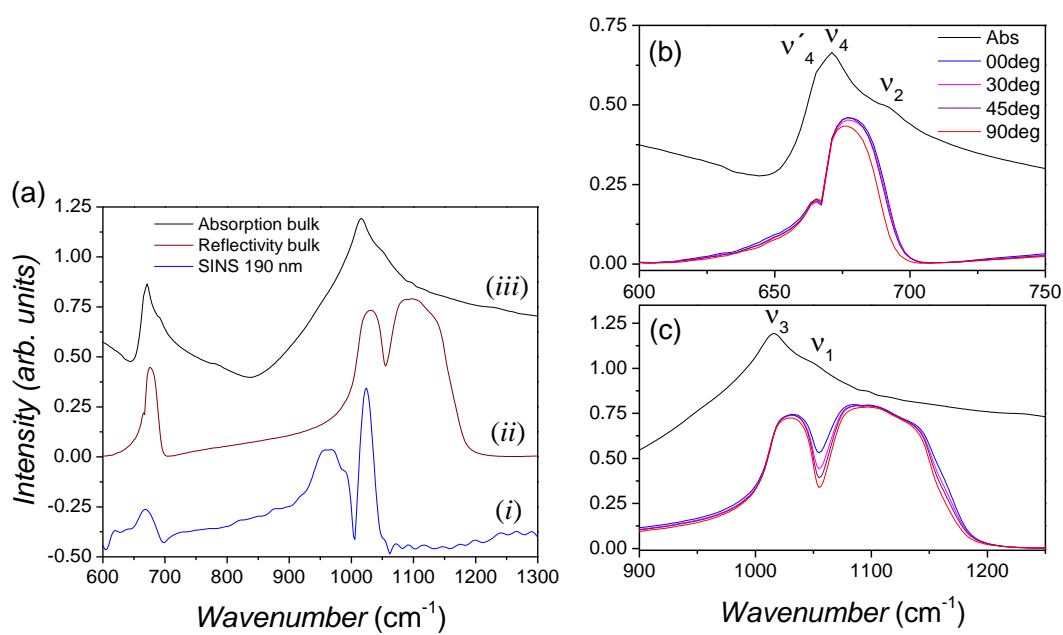

**Figure S2:** (a) SINS (*i*) spectra of nanostructured 190 nm talc/Au compared to unpolarized FTIR (reflectivity, *ii* and absorption, *iii*) spectra of bulk talc samples. (b) and (c) Polarized FTIR reflectivity spectra on the plane of a talc flake, for several angles of the polarizer, and absorption spectra of a powdered sample, in the regions of the SiO$_4$ bending (b) and stretching (c) polar modes.

The SINS spectra of nanostructured talc revealed three main features located at 670 cm$^{-1}$, 960 cm$^{-1}$ and 1025 cm$^{-1}$. In order to check the behavior of these modes in the FTIR spectra of bulk materials, Figure S2 also presents unpolarized absorption (pelletized sample) and polarized reflectivity spectra (oriented crystal flake), in the regions of the SiO$_4$ bending Figure 2S (b) or stretching Figure 2S (c) modes. The reflectivity spectra were taken onto the plane of the crystal flake, and therefore, should reveal directly the in-plane vibrational modes. As shown by the X-ray characterization, talc belongs to a centrosymmetrical triclinic structure (space group P$\bar{1}$, #2). Usually, the polar modes of triclinic crystals have arbitrary polarization directions (not attached to any particular plane or principal direction), and therefore, can be seen for any polarization direction of the infrared electric field. In the case of the P$\bar{1}$ group, all the infrared modes would belong all to the $A_u$ symmetry of the $C_i$ point-group, and, therefore, should be seen in any polarization direction, but with different dielectric strengths (being maximized when the external electric field direction would coincide with the phonon transition moment). This is not what we observe from the polarized spectra of talc. Indeed, the reflectivity spectra of Figures S2 (b) and S2 (c) shows that the phonon modes have essentially the same oscillator strengths, irrespectively of used polarization angle. This result is in complete agreement with polarized absorption measurements on oriented samples done by Farmer[1], who correctly interpreted it in terms of the marked pseudo-hexagonal symmetry ($C_{6v}$) of talc, owing to the hexagonal symmetry of the silicon-oxygen framework. As a consequence, the in-plane infrared spectra of talc looks rather as quasi-degenerate $E_1$ modes of the $C_{6v}$ symmetry, completed by out-of-plane $A_1$ modes of this symmetry (perpendicularly to the flake planes). Based on this approach and comparing absorption spectra of oriented and unoriented samples, Farmer[1] was able to identify the $E_1$ and $A_1$ type modes of talc.

The mid-infrared spectra of talc in the region 600-1300 cm$^{-1}$ shows only SiO$_4$ vibrational modes. In this region, four polar modes are expected for the SiO$_4$ tetrahedron vibration into the $C_{6v}$ symmetry, namely $\nu_1$ and $\nu_3$ stretching modes, and $\nu_2$ and $\nu_4$ bending ones. We indicate the position of these modes in the absorption spectra of Figures S2 (c) and S2 (b), respectively. We found these modes at 1050, 1017, 691 and 671 cm$^{-1}$. Now, by comparing with reflectivity spectra, we can see that modes $\nu_3$ and $\nu_4$ are in-plane modes (inflection points with positive slopes in these spectra), while $\nu_1$ and $\nu_2$ are out-of-plane vibrations (should be absent in the reflectivity spectra). These characteristic features as well as peak positions agree well with those observed by Farmer [1]. However, we were able to discern an extra absorption at around 665 cm$^{-1}$, which also shows up in the reflectivity spectra of Figure S2(b), demonstrating that it should be due to a splitting of the in-plane degenerate $\nu_4$ mode (whose symmetry into the $C_{6v}$ symmetry is $E_1$). Then, we called this mode $\nu'_4$. On the other hand, we were unable to observe a splitting of the degenerate $\nu_3$ mode.

Now, it is important to comment on an important difference between SINS and FTIR reflectivity spectra. Indeed, while both techniques did not discern the presence of the $\nu_2$ are out-of-plane mode, both evidenced the existence of the $\nu_1$ mode (also out-of-plane), but differently. Indeed, SINS shows two independent and strong scattering peaks, while in FTIR spectra this band is denoted by the existence of a dip inside a rather broad band. Note, in Figure S2 (c) that a dip around 1056 cm$^{-1}$ becomes deeper when the polarization angle goes from 0 to 90 degrees. This dip is indeed the signature of the longitudinal optical branch of the $v_1$ mode, and appears because its interaction with the p-polarized light, which increases for higher polarization angles. This p-component exists because the incident infrared light is not exactly normal to the sample´s surface [2].

Finally, once we got enough knowledge about the FTIR response of bulk sample, it becomes clear from peak profiles, oscillator strengths and polarization features that the 670 cm$^{-1}$ band observed by SINS is the in-plane $v_4$ mode. However, due to the large downshift presented by the stretching modes under SINS technique, we are unable to compare their relative positions directly with those obtained by FTIR to attribute the correct positions of $v_1$ and $v_3$ stretching mode of nanostructured talc. A better assignment of the SINS bands required theoretical calculations, as demonstrated in the present paper.

**Section 3 (S3): Optical response of Talc/SiO$_2$ heterostructures**

The optical response of the talc layers (down to mono layers) on top of SiO$_2$ is further explored acquiring SINS spectra. Figure S3 shows SINS spectra of the talc layers atop SiO$_2$. From simple inspection of the Figure S3 and Figure 2 (main text) one can state distinct spectral features of the talc on different substrates. When talc layers are on top of gold film (Figure 2 (a)), only talc-originated modes are observed. However, in Figure S3 it is possible to observe that for thinner talc layers the signal is dominated by the amorphous SiO$_2$ surface phonons at 1120 cm$^{-1}$, while for thicker flakes the intensity of talc modes dominant. Notwithstanding the SiO$_2$ surface phonons are presented, SINS measurements are able to distinguish the monolayer (1 nm thick) talc modes. Note that in both figures, the thicker the talc flake is, the more distinct and intense is the peak, indicating that talc modes show volume dependence. In addition, our results show that the talc flake does not screen the SiO$_2$ substrate and the penetration depth of the near-field optical signal is larger than 90 nm (thickest talc layer measured on SiO$_2$). Finally, due to the combination of stretching and beading modes of talc and SiO$_2$, talc peaks are hardly distinguished from other peaks in the G-talc/SiO$_2$ structure, differently than G-talc/Au as shown in Figure 2 (b) in the main text.

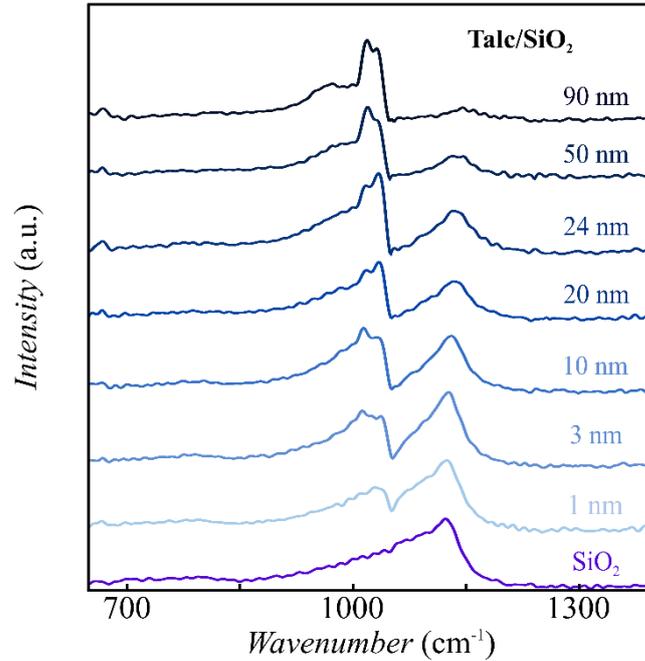

**Figure S3:** SINS spectra of the talc layers on $SiO_2$ substrate, showing the $SiO_2$ substrate surface phonons response at 1120 cm$^{-1}$ and talc modes.

**Section 4 (S4): Graphene-talc plasmon-phonon coupling in $SiO_2$ substrate**

The interaction between graphene and talc was also explored by SINS measurements. In Figure S4 (a) we show SINS spectra measured on the G-Talc/$SiO_2$ sample (positions marked in the inset) in the range between 650 and 1350 cm$^{-1}$. The AFM phase image is seen as inset in Figure S3 (a). For these measurements, we used 20 nm thick talc. The Talc/$SiO_2$ heterostructure (red, solid line in Figure S4 (a)) shows modes at 1120 cm$^{-1}$ (that is assigned to the $SiO_2$ surface phonon) and talc modes at 670 cm$^{-1}$, 960 cm$^{-1}$ and 1025 cm$^{-1}$. For the G/Talc/$SiO_2$ spectrum shown in Figure S4 (a) (black, dotted line), the $SiO_2$ and talc bands are visible and we see a clear enhancement of the peak intensity on the graphene for the G-Talc/$SiO_2$ heterostructure. The net result is that the plasmon of graphene modifies the talc surface phonon response, which is the experimental manifestation of the plasmon-phonon interaction at the graphene-talc interface as discussed in the main text.

The interaction between the graphene and talc flake can be better visualized throughout a linescan. A SINS linescan along the black dotted line marked in the Figure S4(b) with a step size of 50 nm was carried out. To help correlating the feature in the spectra, we marked the observed heterostructure boarders by white dotted lines that extend to Figure S4 (b).

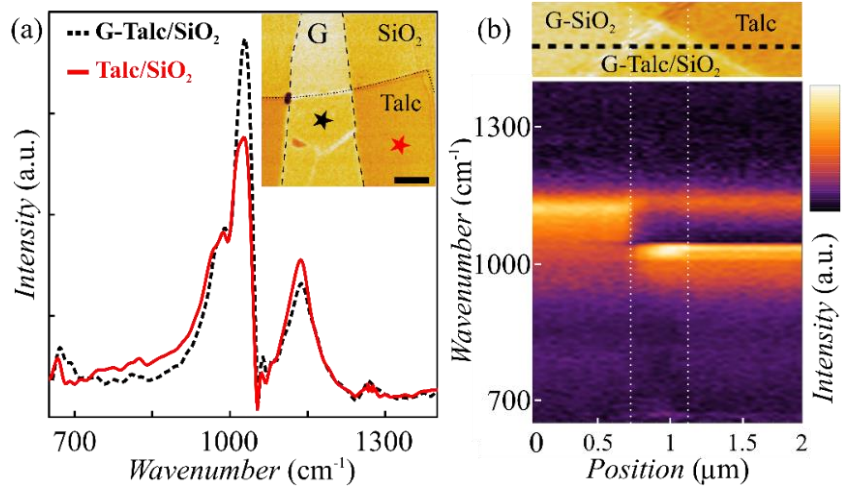

**Figure S4:** (a) SINS spectra of the G/Talc/SiO$_2$ and Talc/SiO$_2$ obtained from the positions marked in the inset. In such measurements, the talc thickness is 24 nm. The scale bar is 1µm. The spectra are dominated by talc modes at 960 cm$^{-1}$ and 1025 cm$^{-1}$. Moreover, a low energy band at 670 cm$^{-1}$ and the SiO$_2$ surface phonons response at 1120 cm$^{-1}$ are also observed. (b) AFM phase image of the region and the linescan correlated. One can see the dependence and intensification of the talc peak at 1025 cm$^{-1}$ in the presence of the graphene layer.

Figure S4 (b) shows the SINS spectra as a function of the sample position. Hereby, the spectral response is color coded as a function of position (x-axis) and wavenumber (y-axis) of the image plot. From the right to the left, we identify the talc modes at 670 cm$^{-1}$, 960 cm$^{-1}$, and at 1025 cm$^{-1}$, and SiO$_2$ band at 1120 cm$^{-1}$. Entering the G/Talc/SiO$_2$ heterostructure, only the mode at 1025 cm$^{-1}$ is enhanced. Reaching the G/SiO$_2$ border, the spectral signature of the talc vanishes and only the SiO$_2$ band is observed. The intensification of such mode enforces our conclusion that the enhancement of this band results from the coupling of the graphene plasmons with the talc phonon.

**Section 5 (S5): Optical phonon characteristics from DFT calculations**

The optical phonon features of talc were obtained by DFT calculations and the results are presented in Tables S1 (bulk) and S2 (monolayer). Frequencies, oscillator strengths, static ($\varepsilon_0$) and electronic ($\varepsilon_\infty$) dielectric permittivities are directly obtained from the calculus. Orientation-averaged dielectric strengths, $\Delta\varepsilon_j$, were calculated from $O.S._j/\Omega_j^2$ and normalized, so that $\Sigma_j\Delta\varepsilon_j = \varepsilon_0 - \varepsilon_\infty$ (the latest constants are average values from the diagonal elements of the respective tensors). Acoustic modes (#1, 2, and 3) were not included in the tables. The modes that have expressive contributions for the observed features (SINS and FTIR) are highlighted. These are modes into the observation region in this work.

**Table S1** – Phonon dispersion parameters for **bulk** talc from the DFT calculations. O.S. are the oscillator strengths [in (D/A)$^2$]. Here, $\varepsilon_0$ = 4.215 and $\varepsilon_\infty$ = 2.546.

| #j-mode | $\Omega_j$ (cm$^{-1}$) | O.S.$_j$ | $10^2\Delta\varepsilon_j$ | #j-mode | $\Omega_j$ (cm$^{-1}$) | O.S.$_j$ | $10^2\Delta\varepsilon_j$ |
|---|---|---|---|---|---|---|---|
| 4 | 97.96 | 0.0669 | 1.26617 | 34 | 443.87 | 0.0179 | 0.0165 |
| 5 | 104.17 | 0.0332 | 0.55567 | 35 | 463.47 | 18.8205 | 15.91292 |
| 6 | 116.17 | 2E-4 | 0.00269 | 36 | 465.6 | 0.2451 | 0.20534 |
| 7 | 126.07 | 1E-4 | 0.00114 | 37 | 477.24 | 8.7494 | 6.97697 |
| 8 | 172.82 | 0.1571 | 0.95532 | 38 | 495.78 | 4E-4 | 2.956E-4 |
| 9 | 183.91 | 0.2018 | 1.08361 | 39 | 505.98 | 0.0035 | 0.00248 |
| 10 | 191.02 | 6E-4 | 0.00299 | 40 | 508.53 | 6.3622 | 4.46824 |
| 11 | 227.52 | 2E-4 | 7.0170E-4 | 41 | 513.61 | 0.0128 | 0.00881 |
| 12 | 229.98 | 0.0293 | 0.10061 | 42 | 658.17 | 0.2057 | 0.08624 |
| 13 | 264.02 | 0.8014 | 2.08803 | **43** | **658.4** | **1.1497** | **0.48169** |
| 14 | 288.12 | 0.064 | 0.14002 | **44** | **668.88** | **7.696** | **3.12414** |
| 15 | 289.09 | 0.0012 | 0.00261 | 45 | 674.93 | 0.13 | 0.05183 |
| 16 | 296.28 | 0.0017 | 0.00352 | **46** | **675.48** | **7.7092** | **3.06864** |
| 17 | 308.11 | 0.4243 | 0.81175 | 47 | 676.83 | 0.0159 | 0.0063 |
| 18 | 325.16 | 0.0021 | 0.00361 | 48 | 756.51 | 0.0358 | 0.01136 |
| 19 | 328.55 | 0.0068 | 0.01144 | 49 | 759.88 | 0 | 0 |
| 20 | 335.38 | 0.5837 | 0.94249 | 50 | 763.23 | 0.0172 | 0.00536 |
| 21 | 337.99 | 9E-4 | 0.00143 | 51 | 766.61 | 0 | 0 |
| 22 | 360.46 | 8E-4 | 0.00112 | 52 | 875.03 | 0.102 | 0.02419 |
| 23 | 362.59 | 0.3609 | 0.49856 | 53 | 875.8 | 0.0054 | 0.00128 |
| 24 | 365.17 | 0.2056 | 0.28002 | **54** | **912.42** | **46.0981** | **10.0567** |
| 25 | 368.98 | 4.8759 | 6.50446 | **55** | **971.98** | **9.9579** | **1.91432** |
| 26 | 373.35 | 0.0183 | 0.02384 | **56** | **972.4** | **52.7741** | **10.13661** |
| 27 | 379.59 | 0.0431 | 0.05433 | **57** | **975.78** | **62.6961** | **11.95911** |
| 28 | 389.22 | 11.8604 | 14.21907 | 58 | 977.61 | 0.055 | 0.01045 |
| 29 | 401.9 | 10.4688 | 11.77126 | 59 | 1015.88 | 1E-4 | 1.760E-5 |
| 30 | 418.1 | 0.006 | 0.00623 | 60 | 1050.07 | 0.0547 | 0.00901 |
| 31 | 428.79 | 17.5114 | 17.29791 | 61 | 1054.31 | 3E-4 | 4.902E-5 |
| 32 | 429.02 | 1.5459 | 1.52542 | 62 | 3720.55 | 2.4574 | 0.03224 |
| 33 | 435.48 | 39.828 | 38.14293 | 63 | 3721.89 | 0.0014 | 1.836E-5 |

**Table S2** – Phonon dispersion parameters for a talc **<u>monolayer</u>** from the DFT calculations. O.S. are the oscillator strengths [in $(D/Å)^2$]. Here, $\varepsilon_0 = 3.672$ and $\varepsilon_\infty = 1.417$.

| #j-mode | $\Omega_j$ (cm$^{-1}$) | O.S.$_j$ | $10^2\Delta\varepsilon_j$ | #j-mode | $\Omega_j$ (cm$^{-1}$) | O.S.$_j$ | $10^2\Delta\varepsilon_j$ |
|---|---|---|---|---|---|---|---|
| 4 | 121.34 | 0.002 | 0.03943 | 34 | 434.64 | 33.4612 | 51.41394 |
| 5 | 128.07 | 0 | 0 | 35 | 436.93 | 25.4178 | 38.64675 |
| 6 | 134.97 | 0.0065 | 0.10357 | 36 | 467.98 | 0.0316 | 0.04188 |
| 7 | 138.93 | 0.0055 | 0.08271 | 37 | 490.28 | 0.8886 | 1.07304 |
| 8 | 174.42 | 0.237 | 2.26128 | 38 | 493.24 | 0.0736 | 0.08781 |
| 9 | 175.09 | 0.3609 | 3.41714 | 39 | 518.57 | 0.0588 | 0.06347 |
| 10 | 225.61 | 6E-4 | 0.00342 | 40 | 526.35 | 0.1259 | 0.13191 |
| 11 | 228.71 | 0.0038 | 0.02109 | 41 | 537.88 | 2.3729 | 2.38071 |
| 12 | 233.46 | 0.0194 | 0.10332 | **42** | **662.68** | **3.6896** | **2.43877** |
| 13 | 274.36 | 0.0397 | 0.15309 | **43** | **671.42** | **2.6131** | **1.68254** |
| 14 | 294.83 | 0.0019 | 0.00634 | 44 | 673.95 | 0.0383 | 0.02448 |
| 15 | 296.76 | 9E-4 | 0.00297 | 45 | 674.75 | 0.0067 | 0.00427 |
| 16 | 300.83 | 3E-4 | 9.6223E-4 | 46 | 685.07 | 0.0019 | 0.00118 |
| 17 | 316.21 | 0.1427 | 0.41426 | **47** | **693.07** | **0.7054** | **0.42627** |
| 18 | 322.41 | 0.0074 | 0.02066 | 48 | 758.72 | 0 | 0 |
| 19 | 324.64 | 0.4052 | 1.116 | 49 | 759.46 | 0.0151 | 0.0076 |
| 20 | 327.79 | 0.0138 | 0.03728 | 50 | 759.53 | 0.0277 | 0.01394 |
| 21 | 339.24 | 0.0021 | 0.0053 | 51 | 760.28 | 9E-4 | 4.5195E-4 |
| 22 | 352.56 | 0.0802 | 0.18729 | 52 | 870.18 | 7E-4 | 2.6834E-4 |
| 23 | 358.53 | 0.2074 | 0.46833 | 53 | 870.83 | 6E-4 | 2.2966E-4 |
| 24 | 364.04 | 0.3686 | 0.80734 | **54** | **969.99** | **57.7233** | **17.80801** |
| 25 | 369.59 | 3.6932 | 7.84804 | **55** | **970.83** | **9.736** | **2.99842** |
| 26 | 371.63 | 0.3184 | 0.66919 | **56** | **975.16** | **65.6784** | **20.04793** |
| 27 | 379.34 | 0.1278 | 0.25779 | 57 | 975.95 | 0.0781 | 0.0238 |
| 28 | 382.86 | 6.7429 | 13.35259 | **58** | **1000.21** | **13.8522** | **4.01916** |
| 29 | 410.33 | 0.0022 | 0.00379 | 59 | 1013.09 | 0.0593 | 0.01677 |
| 30 | 419.94 | 8.2227 | 13.53439 | 60 | 1093.76 | 0.0248 | 0.00602 |
| 31 | 420.85 | 9.9633 | 16.32854 | 61 | 1094.91 | 0.0449 | 0.01087 |
| 32 | 426.56 | 11.1684 | 17.81679 | 62 | 3726.31 | 0.6714 | 0.01404 |
| 33 | 428.93 | 1.9343 | 3.05176 | 63 | 3731.57 | 0.0359 | 7.4836E-4 |

As it can be seen by the very close frequency positions, some of the highlighted modes are quasi-degenerate modes. This are the case for modes 55, 56 and 57 of bulk talk (equivalently 54, 55 and 56 for a monolayer) around 970 cm$^{-1}$, that are indistinguishable experimentally in SINS or FTIR spectra, namely the $\nu_3$ modes. By generating animations and drawings, plotting arrows proportional to the atomic displacements, we see that all these individual modes correspond to in-plane stretching vibrations of Si-O. Since the drawings for a talc monolayer or its bulk are quite similar, we present below (Figure S5) the drawings for modes 54, 55 and 56 of the monolayer.

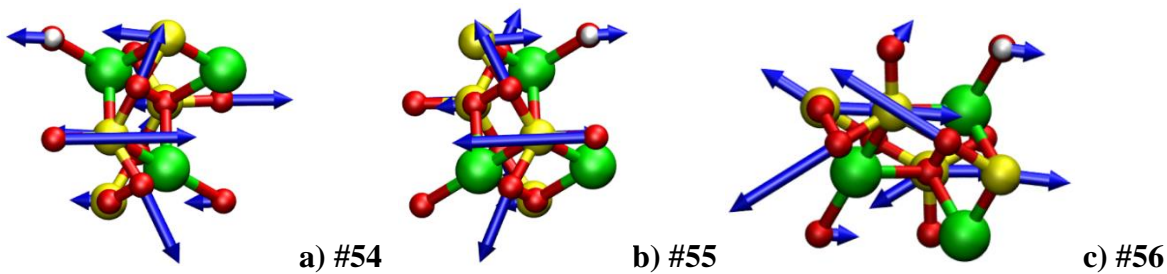

**Figure S5:** The relative amplitudes and directions of motions of atoms in a talc monolayer for the quasi-degenerate ($\nu_3$) modes #54 (calculated at 970.0cm$^{-1}$), #55 (calculated at 970.8cm$^{-1}$) and 56 (calculated at 975.2cm$^{-1}$).

The same analysis was applied to the other highlighted modes. In this way, we could assign the quasi-degenerate 44&46 modes (bulk), or equivalently 42&43 modes (monolayer), to the in-plane $\nu_1$ vibrations around 670 cm$^{-1}$. Our results show that these modes are essentially due to librations of the OH groups, as exemplified for the monolayer, in the Figure S6 below.

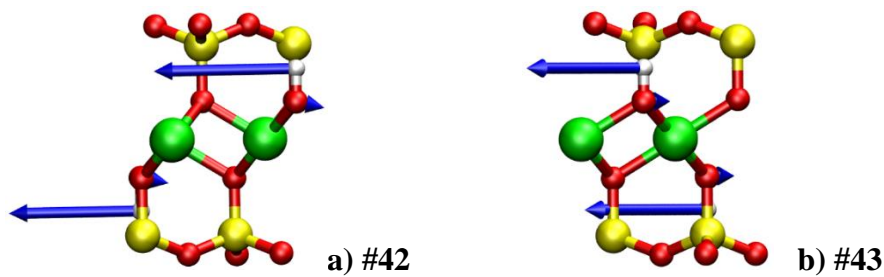

**Figure S6:** The relative amplitudes and directions of motions of atoms in a talc monolayer for the quasi-degenerate in-plane $\nu_1$-type modes #42 (calculated at 662.7 cm$^{-1}$) and 43 (calculated at 671.4 cm$^{-1}$).

The remaining highlighted modes are out-of-plane stretching and deformation vibrations of the SiO$_4$ tetrahedra. For the talc monolayer, these modes are calculated respectively at 1000 cm$^{-1}$ (#58) and 693 cm$^{-1}$ (#47). The out-of-plane 1000 cm$^{-1}$ mode corresponds to our measured $\nu_2$ feature. The lower frequency mode is not seen by SINS, but appears as a faint band in the FTIR absorption, as shown in the next section. For the bulk sample, these out-of-plane modes are calculated to occur in a lower frequency. This can be due to the fact that long-range van-der-Waals forces between the talc layers are neglected in the DFT calculations, influencing the calculations of the out-of-plane modes. Therefore, the calculated out-of-plane modes would have lower frequencies than the observed ones. Thus, the calculus of the out-of-plane modes of a monolayer must be more reliable than for bulk sample. Another consequence is that the frequencies of in-plane modes for monolayer and bulk are quite close, since they are less affected by the interlayer interactions. Below, in Figure S7 we present the sketch of the out-of-plane highlighted modes of talc monolayer.

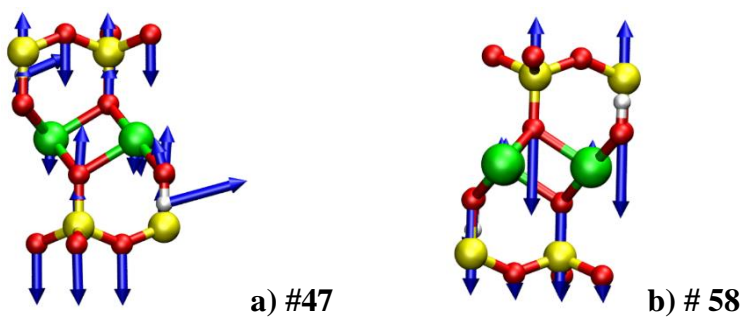

a) #47    b) # 58

**Figure S7:** The relative amplitudes and directions of motions of atoms in a talc monolayer for the out-of-plane modes #47 (calculated at 693 cm$^{-1}$) and #58 (calculated at 1000 cm$^{-1}$). The latest mode was identified as $\nu_2$ in this work.

Finally, we have calculated the 60 infrared-active modes, from which around 30 have enough strength to be candidates to be experimentally discerned according to our theoretical calculations. In the Figure S8 is presented the Im ($\varepsilon$) simulated functions of bulk and talc monolayer in the whole infrared vibrational range, from the DFT calculations. Note, however, that in our SINS experimental window, covering the region 600-1600 cm$^{-1}$, a relatively lower number of modes are active. These are the modes addressed in our work.

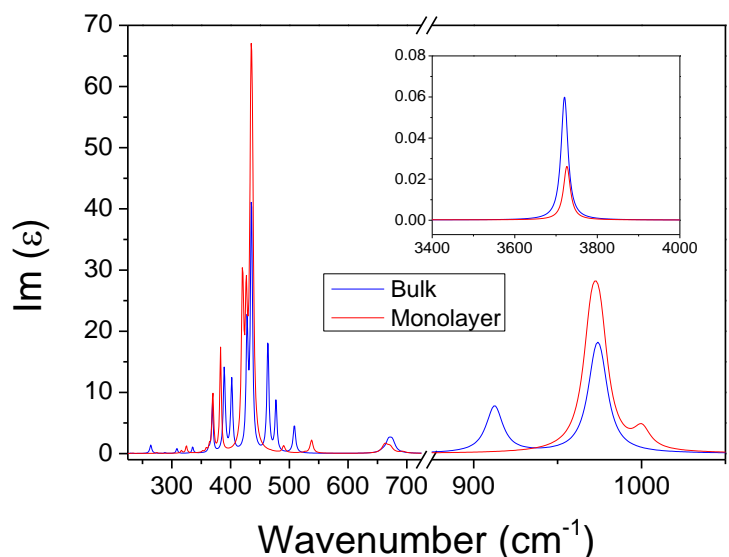

**Figure S8:** Simulated Im ($\varepsilon$) functions of all modes obtained by our DFT calculations for bulk and monolayer talc.